\documentclass[twocolumn,prl,superscriptaddress]{revtex4}
\pdfoutput=1

\usepackage[english]{babel}
\usepackage{amsmath}
\usepackage{graphicx}
\usepackage{amsbsy, amstext}
\usepackage{hyperref}
\usepackage{amsmath, amsthm, amssymb, upref, mathrsfs, amsfonts}
\usepackage{epstopdf}
\usepackage{color}

\newcommand{\be}{\begin{eqnarray}}

\newcommand{\ee}{\end{eqnarray}}

\newcommand{\ds}{\langle \delta_{\gamma}(x)^{2}\rangle_{p}}

\newcommand{\R}{\mathcal{R}}
\newcommand{\zmi}{z_{\mu,i}}
\newcommand{\zmf}{z_{\mu,f}}
\newcommand{\Cl}{C^{\mu T}_{l}}
\newcommand{\fnl}{f_{NL}}

\newcommand{\rs}{r}
\newcommand{\ks}{k_{s}}
\newcommand{\ex}[1]{\langle #1 \rangle}

\begin{document}

\title{A New Window on Primordial non-Gaussianity}

\author{Enrico Pajer}
\affiliation{Department of Physics, Princeton University, Princeton, NJ 08544, USA}
\author{Matias Zaldarriaga}
\affiliation{School of Natural Sciences, Institute for Advanced Study, Princeton, NJ 08540 USA}

\begin{abstract}
We know very little about primordial curvature perturbations on scales smaller than about a Mpc. Measurements of the $\mu$-type distortion of the CMB spectrum provide the unique opportunity to probe these scales over the unexplored range from $50$ to $10^{4}\,{\rm Mpc}^{-1}$. This is a very clean probe, in that it relies only on well-understood linear evolution. Also, just the information about the low multipoles ($l\sim100$) of $\mu$ is necessary. We point out that correlations between $\mu$-distortion and temperature anisotropies can be used to test Gaussianity at these very small scales. In particular the $\mu\-T$ two-point correlation is proportional to the very squeezed limit of the primordial bispectrum and hence measures $f_{NL}^{loc}$, while $\mu\-\mu$ is proportional to the primordial trispectrum and measures $\tau_{NL}$. We present a Fisher matrix forecast of the observational constraints on $f_{NL}^{loc}$ and stress that a cosmic variance limited experiment could in principle reach $\Delta f_{NL}^{loc}\sim \mathcal{O}(10^{-3})$.

\end{abstract}

\maketitle


The initial conditions of our universe can be described in the simplest case just by specifying at some early time the probability distribution function of the adiabatic mode on a wide range of scales. Since then most of these scales have evolved in a complicated and often non-linear way until today. Therefore it is very hard to find accurate probes of these initial conditions, with two notable exceptions: large scale structures today, which still have not entered a fully non-linear regime, and the anisotropies of the Cosmic Microwave Background (CMB) radiation. Because of complicated non-linear dynamics on the one side and Silk damping on the other, both probes are useful only at scales of order a Mpc or larger. Hence it is very interesting to find ways to investigate and constrain scales outside of the $\{10^{-4}-1\}\, {\rm Mpc }^{-1}$ window, which we have explored so far.

In \cite{Hu:1994bz} it was shown that\footnote{For a recent review of the history of the subject as well as a comprehensive bibliography we refer the reader to \cite{Chluba:2011hw}}, due to the dissipation of acoustic waves, the \textit{spectral distortion} of the CMB can be used to constrain the (integrated) primordial power spectrum in the approximate range $50\lesssim k \,{\rm Mpc}\lesssim 10^{4}$. This mechanism provides us with the probably unique chance to probe primordial perturbations that by now have been completely erased by Silk damping and swamped by complicated gravitational dynamics. Although we know that the primordial perturbations in the CMB/LSS window are close to Gaussian and hence well described by just the power spectrum, we know near to nothing about the statistical properties at much smaller scales. 

In this paper we show that, besides probing the integrated power spectrum, the CMB spectral distortion is a potentially powerful tool to constrain deviations from Gaussianity at scales $50\lesssim k \,{\rm Mpc}\lesssim 10^{4}$. Correlations between the $\mu$-type distortion and temperature anisotropies of the CMB provide a direct measurement of the primordial bispectrum (in the squeezed limit), while self-correlations of $\mu$-distortion measure the primordial trispectrum. It is important to stress that this probe is particularly clean in that it only relies on well understood CMB physics and is largely unaffected by gravitational non-linearities.

The rest of the paper is organized as follows. We start by reviewing the relevant aspects of CMB physics. Then, following \cite{Hu:1994bz,Chluba:2011hw,Khatri:2011aj}, we derive an analytic estimate for the $\mu$-distortion which we use to compute two-point self-correlations and cross correlation with temperature anisotropies. Finally we present a Fisher matrix forecast together with some consideration of observational prospects and we conclude with a summary.


\section{CMB distortion}


At early times ($z\gg10^{6}$) the universe is well described by a hot photon-baryon plasma. The number density of photons $n(\nu)$ per frequency interval is given to very high accuracy by the black body spectrum $n(\nu)= \left(e^{x}-1\right)^{-1}$ 
where $x\equiv h\nu/(k_{B}T)$. The equation that describes the subsequent evolution of the photon number density is the full Boltzmann equation (when restricted to Compton scattering this is known as the Kompaneet's equation \cite{Kom}). This equation has three interesting regimes. Before $z\simeq z_{\mu,i}\equiv 2 \times 10^{6}$ any energy released into the photon-baryon plasma is quickly thermalized by elastic and double Compton scattering ($e^{-}\gamma+\rightarrow e^{-}+2\gamma$), which are still very efficient \cite{ref1}. The end result is again a black-body spectrum with now a higher $T$ and a larger total number of photons $N$. After $z_{\mu,i}$ double Compton scattering (as well as bremsstrahlung) becomes less efficient and the total number of photons is approximately frozen \footnote{Strictly speaking this is not true deep in the Rayleigh�Jeans tail ($\nu\rightarrow 0$), but it is a good approximation for the range of frequencies relevant for observations. As we will see shortly this is quantify by the frequency dependence of the chemical potential.}. For $ z_{\mu,f}\lesssim z\lesssim z_{\mu,i}$, with $z_{\mu,f}\equiv 5\times10^{4}$, equilibrium is still achieved after an energy injection due to elastic Compton scattering, but since this process does not change the number of photons the end result is a Bose-Einstein distribution \cite{ref2}, i.e.~$n(\nu)= \left[e^{x+\mu(x)}-1\right]^{-1}$ where $\mu$ is a frequency dependent chemical potential (rescaled by $k_{B}T$ so that it is dimensionless). The Kompaneets equation shows that $\mu(x)$ deviates from a constant only at very low frequencies, i.e.~$\mu(x)=\mu_{0}e^{-x_{c}/x}$, with $x_{c}\simeq 5 \times 10^{-3}$. Henceforth we approximate $\mu$ as constant, which is valid everywhere except deep in the Rayleigh-Jeans tail ($\nu\rightarrow 0$). Finally for $z\lesssim z_{\mu,f}$ even Compton scattering is not efficient enough to establish kinetic equilibrium between matter and radiation. The distortion created after this moment is known as y-type and is relevant e.g.~for the Sunyaev-Zel'dovich effect \cite{ref3}. Of course this is a simplified picture since there is no sharp transition between one regime and the next. For the purpose of analytical estimates we will take the period responsible for the creation of $\mu$-distortion to be $z_{\mu,f}\lesssim z\lesssim z_{\mu,i}$ with the numerical values given above. As we will see, due to a logarithmic dependence on the size of this interval, changing these values by factors of order unity will not alter the main results. It should be clear though that for precise predictions one needs to study the system numerically.

We will be interested in the energy injection coming from the dissipation of acoustic waves of the adiabatic mode (Silk damping) as these re-enter the horizon and start oscillating. Other sources of distortion are present (e.g.~adiabatic cooling \cite{Khatri:2011aj}) and the physics of the system is very rich. Our working assumption here is that either all other sources lead to a smaller and therefore negligible distortion, as it is the case if the primordial power spectrum is not too red tilted, or that all other relevant effects are understood with a high enough precision to be subtracted off leaving the $\mu$-distortion caused by Silk damping as the only signal.


\section{$\mu$-distortion}\label{s:mu}

In this section, following \cite{Hu:1994bz,Chluba:2011hw,Khatri:2011aj}, we derive a formula that relates the late time $\mu$-distortion to the primordial power spectrum. Using the Bose-Einstein distribution plus the fact that the total number of photons is constant, for an amount of energy (density) released into the plasma $\delta E$ one finds that $\mu\simeq 1.4 \delta E/E$. Hence, let us estimate the energy injection due to damping of acoustic waves. The energy density of a density wave is given by\footnote{This formula contains the correct relativistic factor $c_{s}^{2}/(1+c_{s}^{2})$ and for small $c_{s}^{2}$ reduces to the non-relativistic result used e.g.~in \cite{Hu:1994bz}. For $c_{s}^{2}=1/3$ The difference is exactly the factor $3/4$ found in \cite{Chluba:2012gq}.} $Q=\rho  \ds c_{s}^{2}/(1+c_{s}^{2})$, with $c_{s}$ the sound speed, $\rho$ the density and $\delta$ the dimensionless amplitude of oscillations averaged over a period (indicated by $\langle \rangle_{p}$ to differentiate it from the quantum/ensemble average $\langle\rangle$). Since at this time the universe is dominated by radiation we take $\rho=\rho_{\gamma}$ and $c_{s}^{2}\simeq1/3$. Then one has
\be 
\frac{\delta E}{E}\simeq -\int_{\zmi}^{\zmf}\frac{d}{dz}\frac{Q}{\rho_{\gamma}}\simeq\frac{1}{4}\ds |_{\zmf}^{\zmi}
\ee

\begin{figure}
\includegraphics[width=.475\textwidth]{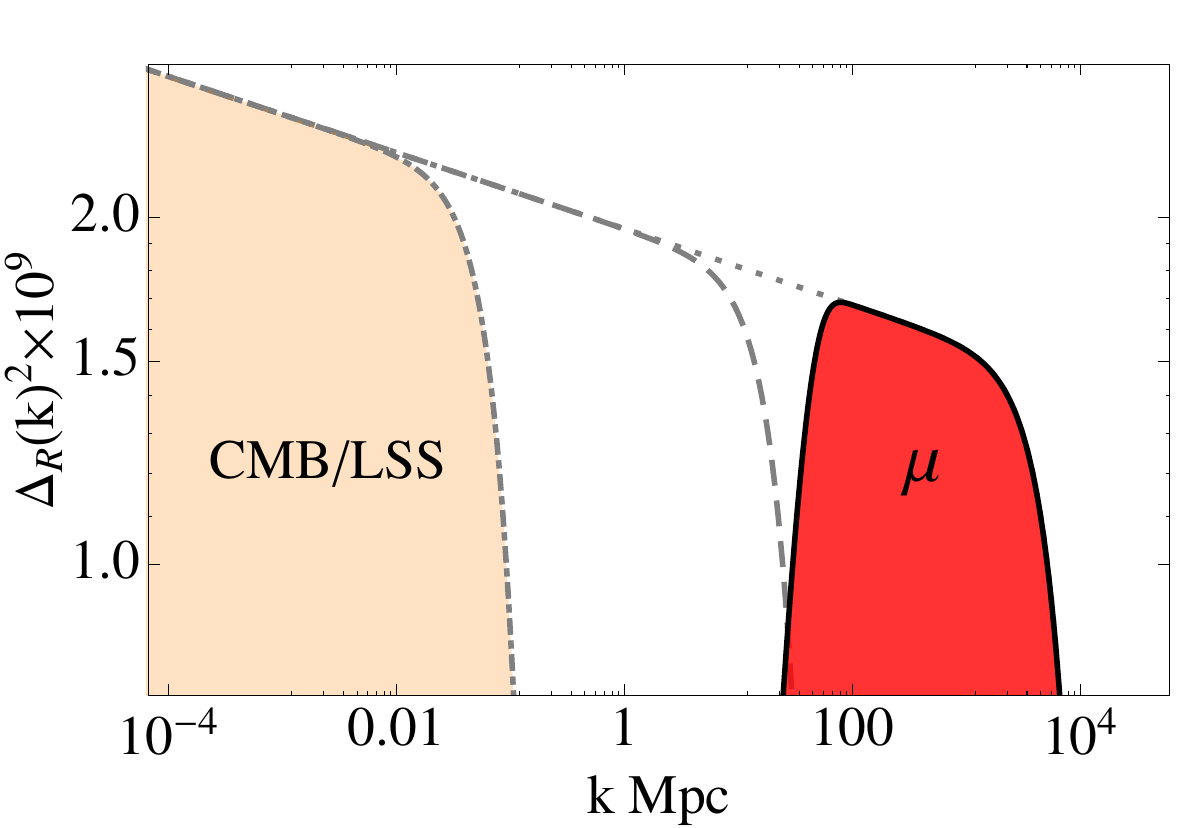}
\caption{The figure shows the power spectrum with Silk damping as function of $\log k$. The dotted, dashed and dot-dashed lines are $\Delta_{R}^{2}e^{-2k^{2}/k_{D}^{2}}$ at $z_{\mu,i}=2\times 10^{6}$, $z_{\mu,f}=5\times 10^{4}$ and $z_{L}=1100$ respectively. The red area on the right indicated by $\mu$ is the difference of the power spectrum between $z_{\mu,i}$ and $z_{\mu,f}$. Once integrated over $\log k$ this gives the $\mu$-distortion. For comparison on the left we have highlighted the scales probed by LSS and CMB anisotropies.  \label{f1}}
\end{figure}

We can use the transfer function (see e.g.~\cite{Weinberg:2008zzc})
\be\label{Delta}
\Delta_{\gamma}(k)\simeq 3 \cos \left(k \rs \right) e^{-k^{2}/k_{D}^{2}}\,,
\ee
where, using that $R\equiv 3\rho_{B}/4\rho_{\gamma}\ll1$, the diffusion damping scale is
\be
k_{D}&\equiv& \left[\int_{z}^{\infty}dz\frac{1+z}{6Hn_{e}\sigma_{T}(1+R)} \left(\frac{R^{2}}{1+R}+\frac{16}{15}\right)\right]^{-1/2}\nonumber\\
&\simeq&(1+z)^{3/2} \,4.1\times 10^{-6}\,{\rm Mpc}^{-1}\,,
\ee
and 
\be\label{winside}
k \rs= \int_{0}^{t}\frac{k\,dt}{a\sqrt{3(1+R)}}\simeq \frac{2 k t}{a\sqrt{3}}\,.
\ee
We then have 
\be
\ds&=& 1.45\int \frac{d^{3}k_{1}d^{3}k_{2}}{(2\pi)^{6}}\,\R(k_{1})\R(k_{2})\\
\nonumber && \times\langle\Delta_{\gamma}(k_{1})\Delta_{\gamma}(k_{2})\rangle_{p}e^{i \left(\vec k_{1}+\vec k_{2}\right)\cdot \vec x}\,,
\ee
where $\R$ describes curvature perturbations. Finally to account for the fact that $\mu$ arises from a thermalization process, we use a top-hat filter in real space $W(x)$, which smears the dissipated energy over a volume of radius $\ks^{-1}\gtrsim k_{D,f}^{-1}$.

Summarizing, the deformation parameter $\mu$ is related to primordial perturbations by
\be
\mu(x) &\simeq & 4.6 \int \frac{d^{3}k_{1}d^{3}k_{2}}{(2\pi)^{6}}\R(\vec k_{1})\R(\vec k_{2})e^{i \vec k_{+}\cdot \vec x} W  \left(\frac{k_{+}}{\ks}\right) \\
&& \times \langle\cos \left(k_{1}\rs\right) \cos \left(k_{2}\rs\right)\rangle_{p} \left[e^{-(k_{1}^{2}+k_{2}^{2})/k_{D}^{2}}\right]^{\zmi}_{\zmf} \nonumber
\ee
where $W(k)\equiv 3k^{-3} \left[\sin(k)-k\cos(k)\right]$ is the Fourier transform of the top-hat filter $W(x)$ and $\vec k_{\pm}\equiv \vec k_{1}\pm\vec k_{2}$.
The quantum/ensemble average of $\mu(x)$ gives the log-integral of the primordial power spectrum from $k_{D}(\zmi)\simeq 1.1\times 10^{4}\,{\rm Mpc}^{-1}$ to $k_{D}(\zmf)\simeq 46\, {\rm Mpc}^{-1}$
\be\label{logint}
\langle \mu(x)\rangle\simeq 2.3 \int d\log k \, \Delta_{\R}^{2}(k) \left[e^{-2 k^{2}/k_{D}^{2}}\right]^{i}_{f}\,,
\ee
where
\be
\langle \R^{2} \rangle&\equiv& (2\pi)^{3}\delta^{3}\left(\vec K_{tot}\right) P_{\R}(k)\,,\nonumber \\
P_{\R}(k)&\equiv&\frac{2\pi^{2}\Delta_{R}^{2}(k)}{k^{3}}\,.
\ee
To visualize this, in figure \ref{f1} we plot $\Delta_{\R}^{2}(k) e^{-2k^{2}/k_{D}^{2}}$ for $k_{D}(\zmi)$ (dotted line) and $k_{D}(\zmf)$ (dashed line) as well as their difference (red-filled region on the right), which quantifies the amount of dissipated energy and hence $\mu$-distortion.


\section{Two-point correlation functions}\label{s:2pts}

Let us now compute the two point correlation between a direction dependent $\mu$-distortion $\mu(\hat n)$ and temperature anisotropy $\Delta T(\hat n)$. To make contact with the way observations are analyzed, we will decompose both signals in spherical harmonics, which are an orthonormal basis of functions on the sphere. We start with 
\be
a^{T}_{lm}&\equiv& \int d\hat n \frac{\Delta T (\hat n)}{T}Y^{\ast}_{lm}(\hat n)\nonumber \\
&=&4\pi \frac{3}{5} (-i)^{l}\int \frac{d^{3}k}{(2\pi)^{3}}\R(\vec k)\Delta_{l}(k)Y^{\ast}_{lm}(\hat k)\,,\label{aT}
\ee
where $\Delta_{l}(k)$ is the radiation transfer function. Throughout this paper we will use the Sachs-Wolfe approximation $\Delta_{l}(k)\simeq j_{l}(k r_{L})/3$ with $r_{L}\simeq 14\,$Gpc the distance from last scattering and $j_{l}$ a spherical Bessel function. This is in principle valid only in the range $10\lesssim l \lesssim50$, but for the purpose of the Fisher forecast it will be a reasonable estimate also for higher l's. Let us define in general
\be 
\langle (a^{i}_{lm})^{\ast}a^{j}_{l'm'}\rangle=\delta_{ll'}\delta_{mm'}C_{l}^{ij}\,,
\ee
then the $TT$ correlation gives the well known result
\be
C_{l}^{TT}=\frac{2 \pi}{25} \frac{\Delta_{\R}(k_{p})^{2}}{l(l+1)}\simeq \frac{6.0 \times 10^{-10}}{l(l+1)}
\ee
where $k_{p}\equiv 0.002\,{\rm Mpc}^{-1}$ is a common pivot scale at which $\Delta_{\R}^{2}(k_{p})=2.4\times 10^{-9}$ \cite{Komatsu:2010fb}. The $\mu$-distortion can also be expanded as
\be \label{alm}
a^{\mu}_{lm}&=&4.6 \int d\hat nY^{\ast}_{lm}(\hat n)\frac{d^{3}k_{1}d^{3}k_{2}}{(2\pi)^{6}}\R(\vec k_{1})\R(\vec k_{2})e^{i \vec k_{+}\cdot \vec x}  \\
&& W  \left(\frac{k_{+}}{\ks}\right)\langle\cos \left(k_{1}\rs\right) \cos \left(k_{2}\rs\right)\rangle_{p} \left[e^{-(k_{1}^{2}+k_{2}^{2})/k_{D}^{2}}\right]^{i}_{f}\nonumber\,.
\ee
Using the identities
\be
e^{i\vec k \cdot \vec x}&=&\sum_{l}(2l+1)i^{l}P_{l}(\hat k\cdot \hat x)j_{l}(kx)\,,\\
P_{l}(\hat k\cdot \hat x)&=&\frac{4\pi}{2l+1}\sum_{m=-l}^{l} Y_{lm}(\hat k)Y^{\ast}_{lm}(\hat x)\,,
\ee
one can conveniently rewrite \eqref{alm} as
\be 
&&18.4\pi (-i)^{l}\int\frac{d^{3}k_{1}d^{3}k_{3}}{(2\pi)^{6}}Y^{\ast}_{lm}(\hat k_{+}) \R(\vec k_{1}) \R(\vec k_{2}) W  \left(\frac{k_{+}}{\ks}\right)\nonumber \\
&&\,j_{l}(k_{+}r_{L}) \langle\cos \left(k_{1}\rs\right) \cos \left(k_{2}\rs\right)\rangle_{p} \left[e^{-(k_{1}^{2}+k_{2}^{2})/k_{D}^{2}}\right]^{i}_{f},
\ee
The $T\mu$ correlation is found to be
\be
\Cl&=&6.1\pi\,\frac{9}{25}f_{NL}b\,   \frac{\Delta_{\R}^{4}(k_{p})}{l(l+1)}\ln \left(\frac{k_{D,i}}{k_{D,f}}\right)\nonumber\\
&\simeq&\frac{2.2\times 10^{-16}}{l(l+1)} f_{NL} b \,,
\ee
where we used the local bispectrum\footnote{To simplify our notation we omit the label $loc$ in $f_{NL}^{loc}$. Since local non-Gaussianity is the only form of non-Gaussianity we consider in this work, this should not lead to any confusion.}
\be
\langle \R^{3}\rangle &=&(2\pi)^{3} \delta^{3} \left(\vec K_{tot}\right) \left(-\frac{6}{5} f_{NL}\right) \label{B}\\
&& \times \left[P_{\R}(k_{1})P_{\R}(k_{2})+{\rm 2\,perm's}\right]\nonumber
\ee
and defined
\be 
 \frac{b}{l(l+1)}\equiv \frac{2}{\log \left(\frac{k_{D,i}}{k_{D,f}}\right)} \int d\log k_{+} j_{l}(k_{+}r_{L})^{2}\nonumber \\
 \times \int d\log k_{-} \frac{\Delta_{\R}^{2}(k_{-}/2)}{\Delta_{R}^{2}(k_{p})} \frac{\Delta_{\R}^{2}(k_{+})}{\Delta_{R}^{2}(k_{p})} \nonumber \\
\times  \left[e^{-(k_{+}^{2}+k_{-}^{2})/(2k_{D}^{2})}\right]^{i}_{f}  W  \left(\frac{k_{+}}{\ks}\right)\,, \label{b}
\ee
This formula can be simplified if one assumes a weak scale dependence. Taking $\Delta_{\R}^2(k)=\Delta_{\R}^2(k_p)(k/k_p)^{n_s-1}$ one finds
\be
b\simeq \frac{\left[\left(\frac{k_D}{2k_p}\right)^{n_s-1}\right]^i_f}{(n_s-1)\log\left(\frac{k_{D,i}}{k_{D,i}}\right)}\simeq1+\frac{n_s-1}{2}\log\left(\frac{k_{D,i}k_{D,f}}{4k_p^2}\right)\,,\nonumber
\ee
so that $b\simeq1+12(n_s-1)$ close to the scale invariant limit $n_s\rightarrow 1$.

The $\mu\mu$ self-correlation has a Gaussian and a non-Gaussian contribution. For the former one finds
\be \label{cmm}
C_{l,{\rm Gauss}}^{\mu\mu}&\sim& 3.5\times 10^{-17}\, \frac{\Delta_{\R}^{4}(k_{D,f})}{\Delta_{\R}^{4}(k_{p})} \frac{\ks r_{L}^{-2}}{k_{D,f}^{3}}\nonumber \\
&\lesssim&9\times 10^{-29} 
\ee
where in the last line we took $\Delta_{R}(k_{D,f})=\Delta_{R}(k_{p})$ and in order to get an upper bound $\ks=k_{D,f}$, which is the shortest scale that contributes sizably to dissipation. This scaling can be understood as follows. For a Gaussian field, consider the power spectrum (i.e.~$\mu$-distortion) at small scales $k_{D,f}$. Its fluctuations in regions much more distant than $k_{D,f}^{-1}$ are completely independent. Hence, the expansion of these power fluctuations in multiples is independent of $l$, i.e.~just white noise, for low multiples $l/r_{L}\ll k_{D,f}$. Now let us consider the effect of smearing. Measuring the low $\mu$-distortion multiples, we are effectively averaging over many small-scale (again of order $k_{D,f}$) independent realizations, and hence we are suppressing the individual variance by $N^{-1/2}$. The number of $k_{D,f}$ modes in a spherical shell around the last scattering surface of thickness $k_{s}$ is given by $N\sim k_{D,f}^{3}/(k_{s} r_{L}^{-2})$.


Using the local trispectrum
\be
\langle \R^{4}\rangle &=&(2\pi)^{3} \delta^{3} \left(\vec K_{tot}\right) \tau_{NL} \nonumber\\
&& \times \left[P_{\R}(k_{1})P_{\R}(k_{2})P_{\R}(|\vec k_{1}+\vec k_{3}|)+{\rm 11\,perm's}\right]\nonumber
\ee
for the non-Gaussian contribution one finds
\be\label{cmmng}
C_{l,NG}^{\mu\mu}&\sim&42 \pi \frac{\Delta_{\R}(k_{p})^{6}}{l(l+1)}\tau_{NL}\,b'\ln^{2} \left(\frac{k_{D,f}}{k_{D,i}}\right)\nonumber \\
&\simeq&5.3 \times 10^{-23}\, \tau_{NL}\,\frac{b'}{l(l+1)}\,,
\ee
where $b'$ is defined by
\be
\frac{b'}{l(l+1)}&=&\frac{2}{\ln^{2} \left(\frac{k_{D,f}}{k_{D,i}}\right)}\int d\ln k_{+}d\ln k_{-} d\ln k_{3} \nonumber\\
&& \times \frac{\Delta_{\R}(k_{-}/2)^{2}\Delta_{\R}(k_{+})^{2}\Delta_{\R}(k_3)^{2}}{\Delta_{\R}(k_{p})^{6}} \nonumber\\
&&\times \left[e^{-k_{-}^{2}/(2k_{D}^{2})}\right]^{i}_{f}\left[e^{-2k_{3}^{2}/k_{D}^{2}}\right]^{i}_{f}
\ee
such that $b'\simeq1$ for a scale invariant power spectrum. Notice that, for a given amount of primordial non-Gaussianity, the ratio of the non-Gaussian to Gaussian contributions to the two-point self correlation is much larger for $\mu$ than it is for $T$. In other words, $C_{l}^{\mu\mu}$ is more sensitive in relative terms to non-Gaussianity than $C_{l}^{TT}$.

To discuss the practical observability of the above signals we need to consider experimental noise. For $T$ this is negligible since we will only consider low multiples ($l\lesssim \mathcal{O}(100)$). To model the noise for $\mu$ we assume a Gaussian beam and use \cite{Dodelson:2003ft}
\be
C_{l}^{\mu\mu,N}\simeq w_{\mu}^{-1}\,e^{l^{2}/l_{\rm max}^{2}}\,,
\ee
where $w_{\mu}$ is the sensitivity to $\mu$ and $l_{\rm max}$ is related to the beam size of the experiment. For example, for an experiment like PIXIE \cite{Kogut:2011xw} the beam size is $\theta_{FWHM}=1^{\circ}.6$ which leads to $l_{max}\simeq 84$ and the sensitivity is $w_{\mu}^{-1/2}\simeq \sqrt{4\pi}\times 10^{-8}$. So a figure of merit to keep in mind is 
\be\label{cmmn}
C_{l}^{\mu\mu,N}\simeq 1.3 \times 10^{-15}\,e^{l^{2}/l_{max}^{2}}\,.
\ee


\section{Fisher matrix analysis}\label{s:fm}

We are now going to use the above results to perform a Fisher forecast for the bounds that $\mu$-distortion can put on $f_{NL}$ local. The fact that we invoked the Sachs-Wolfe approximation instead of using numerical transfer functions should not alter much the estimate of the signal to noise ratio because the acoustic peaks in $T$ mostly scale away.
The $1\times1$ Fisher matrix
\be
F\equiv-\frac{\partial^{2}}{\partial\fnl^{2}}\ln L=\frac{\partial^{2}}{\partial\fnl^{2}}\frac{\chi^{2}}{2}=\sum_{l}\frac{\Cl\,\Cl}{\sigma^{2}_{l}}\,,\nonumber
\ee
gives us the signal-to-noise ratio via $S/N=f_{NL}\sqrt{F}$. For mild non-Gaussianity, $\fnl\lesssim10^{5}$, we can estimate the noise in the measurement of each $l$ as 
\be
\sigma_{l}^{2}&=&\langle (\Cl)^{2}\rangle-\langle \Cl\rangle^{2}\\
&\simeq&\frac{1}{2l+1} C_{l}^{TT} C_{l}^{\mu\mu,N}\,,\nonumber
\ee
where we used that $\mu$ and $T$ instrumental noises are uncorrelated, $C_{l}^{TT}\gg  C_{l}^{TT,N}$ and $C_{l}^{\mu\mu}\ll C_{l}^{\mu\mu,N}$ as seen by comparing \eqref{cmm} with \eqref{cmmn}. Then the signal-to-noise ratio can be written as
\be\label{ref}
\frac{S}{N}\simeq \frac{12}{5} f_{NL} \Delta_{R} \sqrt{\log \left(\frac{l_{\rm max}}{2}\right)}\left(\frac{S}{N}\right)_{\mu}
\ee
where $(S/N)_{\mu}$ is the signal-to-noise ratio for the $\mu$-distortion averaged over the whole sky. Apart from the numerical coefficient, this relation is to be expected. The long-scale fluctuations of $\mu$ around its average that correlate with temperature fluctuations are generated by the non-Gaussian part of $\R$ and hence are suppressed with respect to the $\mu$ monopole by $f_{NL}\R$. Notice that reducing the beam size of the experiment improves the signal to noise ratio only logarithmically. Plugging in numbers one finds the figure of merit
\be
\frac{S}{N}&\simeq &0.7\times 10^{-3}\,b\,f_{NL}\, \left(\frac{\sqrt{4\pi}\times 10^{-8}}{w_{\mu}^{-1/2}}\right)\,.
\ee
where $\Delta \mu=10^{-8}$ is the estimated one sigma error on the $\mu$-distortion monopole of an experiment like PIXIE and $b$, defined in \eqref{b}, is of order one if the primordial power spectrum is approximately scale invariant. 

There are at least two classes of models in which the $\mu T$ correlation could provide the strongest constraints on non-Gaussianity already with PIXIE's sensitivity. First, models in which the power spectrum grows at small scales \cite{largeb}, since then $b\gg1$. In most of these models $b$ can be as large as $\mathcal{O}(10^{2})$ leading to $\Delta f_{NL}^{loc}\sim \mathcal{O}(10)$ for PIXIE. Second, models in which the bispectrum diverges faster than the local template in the squeezed limit, i.e.~$\ex{\R^{3}} \propto k_{3}^{-\alpha}$ for $k_{3}\rightarrow 0$ with $\alpha>3$ (see e.g.~\cite{superlocal} for a phenomenological model). 

Before concluding, let us remark that while instrumental noise can be improved, there is a lower bound for the noise imposed by nature, often referred to as cosmic variance. In this respect the $\mu T$ cross correlations possess an important advantage as compared with the temperature bispectrum, which is another probe of the primordial local bispectrum\footnote{Remember though that the two observables constrain different scales and are therefore complementary rather than competing.}. Cosmic variance, and more generally sampling variance, scales with the number of modes $N$ as $N^{-1/2}$. For a primordial bispectrum of the local type it is useful to distinguish between long and short modes. The latter are the same for both observables, provided one measures the low multiples of $\mu$ ($l\sim 100$). On the other hand, the number of short modes in the temperature bispectrum is about $l_{\rm max}^{2}$ and due to Silk damping $l_{\rm max}$ can never be larger than a few thousand. In $\mu T$ cross correlations the number of short modes is much larger and, as we said, can be estimated as $k_{D,f}^{3}/(\ks r_{L}^{-2})\gtrsim 10^{12}$. This means that, for comic variance limited experiments, the bispectrum's sensitivity is $\Delta f_{NL} \sim \mathcal{O}(5)$, while using $\mu T$ correlation one can in principle reach $\Delta f_{NL} \sim \mathcal{O}(10^{-3})$.


\section{Conclusions}\label{s:c}

Measurements of the $\mu$-distortion of the CMB spectrum offer the unique opportunity to probe primordial perturbations in the otherwise inaccessible range $50\lesssim k \,{\rm Mpc} \lesssim 10^{4}$. We have shown that the two-point self correlations of $\mu$-distortion anisotropies and cross correlations with temperature anisotropies provide a direct measurement of the primordial tri- and bispectrum in the squeezed limit, respectively. We have performed a Fisher matrix forecast for the bounds on $f_{NL}$ local and found that a bound of $|\Delta f_{NL}| \lesssim10^{3}$ is achievable already with current technology \cite{Kogut:2011xw}. It is worth stressing again that in the case of mild non-Gaussianity the monopole $\mu$-distortion is expected to be larger than the anisotropic part (see \eqref{ref} for a quantitative statement).

 It would be interesting to further study the performances of a dedicated experiment. We should stress that one should be careful in comparing this bound with those from CMB/LSS, since it applies to very small scales, around a kpc, which are completely unconstrained so far. 

We have also noticed that the $\mu T$ cross correlation is contaminated by a much smaller cosmic variance than the temperature bispectrum \footnote{It would be interesting to check whether any astrophysical foreground can contaminate the signal.} and for an ideal experiment it could probe an $f_{NL}$ much smaller than unity. In this respect it is important to stress that for all practical purposes single-field inflation predicts a primordial bispectrum that vanishes in the squeezed limit. The well known consistency relation \cite{Maldacena:2002vr} $f_{NL}=n_{s}-1$ is derived using comoving momenta. When comparing with observations one should use physical momenta instead. Doing this leads to an additional term that exactly cancels the $n_{s}-1$. As long as one considers other sources of primordial local non-Gaussianity so that $f_{NL}\gg n_{s}-1$, this slow-roll correction can be safely neglected as we have done in the present derivation.

Finally, given the pace with which observations have improved in recent years, there is good reason to hope that CMB spectral distortion will one day tell us another little bit about the very early stages of our universe.


%
%
%
%


\section*{Acknowledgments}

Is a pleasure to thank A.~Kusaka, D.~Meerburg, J.~Peebles and D.~Spergel for very useful discussions. E.~P.~is supported in part by the Department of Energy grant DE-FG02-91ER-40671. M.~Z.~is supported in part by the National Science Foundation grants PHY-0855425,
AST-0506556 and AST-0907969, and by the David \& Lucile Packard and the John D. \& Catherine
T. MacArthur Foundations.



\begin{thebibliography}{99}



\bibitem{Hu:1994bz}
W.~Hu, D.~Scott and J.~Silk,
Astrophys.\ J.\ {\bf 430} (1994) L5
[arXiv:astro-ph/9402045].

\bibitem{Khatri:2011aj}
R.~Khatri, R.~A.~Sunyaev and J.~Chluba,
arXiv:1110.0475 [astro-ph.CO].


\bibitem{Chluba:2011hw}
J.~Chluba and R.~A.~Sunyaev,
arXiv:1109.6552 [astro-ph.CO].



\bibitem{Kom}
A.~S.~Kompaneets, Zh.~Eksp.~Teor.~Fiz.~31, 876 [Sov.~Phys.~JETP 4, 730 (1957)]

\bibitem{ref1}
Danese L., de Zotti G., 1982, AA, 107, 39 Hu W., Silk J., 1993, Phys.
Rev. D, 48, 485

\bibitem{ref2}
Sunyaev R. A., Zeldovich Y. B., 1970c, ApSS, 7, 20 Illarionov A. F.,
Sunyaev R. A., 1975a, SvA, 18, 413

\bibitem{ref3}
Zeldovich Y. B., Sunyaev R. A., 1969, ApSS, 4, 301

\bibitem{Komatsu:2010fb}
E.~Komatsu {\it et al.} [WMAP Collaboration],
Astrophys.\ J.\ Suppl.\ {\bf 192} (2011) 18
[arXiv:1001.4538 [astro-ph.CO]].


\bibitem{Weinberg:2008zzc}
S.~Weinberg,
{\it Oxford, UK: Oxford Univ. Pr. (2008) 593 p}




\bibitem{ddz}
L.~Danese and G.~de Zotti,
Nuovo Cimento (1971-1977)
Volume 7, Number 3, 277-362, DOI: 10.1007/BF02747276,
J.D.~Barrow and P. Coles
Mon. Not. Roy. astr. Soc., 248, 52-57 (1991) 





\bibitem{Dodelson:2003ft}
S.~Dodelson,
 Amsterdam, Netherlands: Academic Pr. (2003) 440 p
 
 
\bibitem{Chluba:2012gq}
J.~Chluba, R.~Khatri and R.~A.~Sunyaev,
arXiv:1202.0057 [astro-ph.CO].




\bibitem{largeb}
See e.g.~J.~Chluba, A.~L.~Erickcek and I.~Ben-Dayan,
arXiv:1203.2681 [astro-ph.CO], and references therein.


\bibitem{superlocal}
J.~Ganc,
Phys.\ Rev.\ D {\bf 84} (2011) 063514
[arXiv:1104.0244 [astro-ph.CO]].



\bibitem{Kogut:2011xw}
A.~Kogut {\it et al.},
JCAP {\bf 1107} (2011) 025
[arXiv:1105.2044 [astro-ph.CO]].


\bibitem{Maldacena:2002vr}
J.~M.~Maldacena,
JHEP {\bf 0305} (2003) 013
[arXiv:astro-ph/0210603].




\end{thebibliography}
\end{document}